\documentclass[aps,prl,twocolumn,nopacs,superscriptaddress,longbibliography]{revtex4-1}

\usepackage[breaklinks=true,colorlinks,citecolor=blue,linkcolor=blue,urlcolor=blue]{hyperref}
\usepackage{epsfig,mathrsfs,color,latexsym,subfigure,marginnote,graphicx,verbatim,relsize,mathrsfs,color,array,amsmath,amsfonts,amssymb,graphicx}
\begin{document}

\title{Evolution of high-order Van Hove singularities away from cuprate-like band dispersions and its implications for cuprate superconductivity}

\author{Robert S. Markiewicz}
\email{r.markiewicz@northeastern.edu}
\affiliation{Department of Physics, Northeastern University, Boston, Massachusetts 02115, USA}

\author{Bahadur Singh}
\email{bahadur.singh@tifr.res.in}
\affiliation{Department of Condensed Matter Physics and Materials Science, Tata Institute of Fundamental Research, Colaba, Mumbai 400005, India}

\author{Christopher Lane}
\affiliation{Theoretical Division, Los Alamos National Laboratory, Los Alamos, New Mexico 87545, USA}
\affiliation{Center for Integrated Nanotechnologies, Los Alamos National Laboratory, Los Alamos, New Mexico 87545, USA}

\author{Arun Bansil}
\affiliation{Department of Physics, Northeastern University, Boston, Massachusetts 02115, USA}

\begin{abstract}

We discuss the evolution of high-order Van Hove singularities (hoVHSs) that carry faster-than logarithmic
divergences over a wide range of parameters in cuprate-like electronic band dispersions.
Numerical analysis gives insight into the quantization of the VHS power-law-exponent $p_V$ and into
transitions between hoVHSs with different values of $p_V$. The cuprates are found to lie in the
parameter regime where the amplitude of the hoVHS is not too large. Our study indicates that
the occurrence of high-temperature superconductivity requires simultaneous tuning of two different
competing orders (antiferromagnetism and the density wave associated with the hoVHS in cuprates),
which is why it is so rare.

\end{abstract} 

\maketitle

{\it Introduction.$-$}  Two-dimensional (2D) saddle-point Van Hove singularities (VHSs)\cite{VHS} have attracted theoretical interest because a diverging electronic density of states (DOS) can drive an instability even for an arbitrarily small interaction strength. Regarding the cuprates, even though the VHS lies close to the Fermi level in optimally-doped La-cuprates, this is not the case in other high-$T_c$’s, and the associated weak logarithmic divergence of the DOS often results in only low-temperature instabilities. The divergence in VHSs can be much stronger, even power law, in 2D and 3D materials \cite{Liang, Liang3, MBMB, Marsig}. Interest in VHSs has revived recently following the discovery that they are highly tunable and correlate with exotic phases in twisted-bilayer graphene, TMDs, and other materials \cite{Liang,Liang_TMD,Tgraphene,SRO327}, although the modeling of twisted systems with small twist angles presents challenges due to the large unit cell involved.

We expect 2D and 3D single phase materials to also host hoVHSs. Since the DOS divergence in hoVHSs supports quantized power-law exponents, the problem is well suited for machine-learning approaches based on correlating a specific property -- such as superconductivity -- for a given power-law divergence in a large number of materials. However, a recent machine-learning study found little correlation between superconductivity and DOS peaks for square lattice materials\cite{CavaEmptor}.

In this paper, we start with the cuprates, and explore the parameter space obtained by allowing the hopping parameters to vary over a broad range of values. Over the full 2D parameter range we considered, all hoVHSs fall on a single line, and all except one have {\it the same} power-law exponent with an amplitude that varies systematically.  Remarkably, although the cuprates share the same exponent, they have one of the smallest amplitudes, thus confirming the machine-learning result. We show how a more nuanced study can hone in on the parameter range relevant to cuprates.

{\it High-order VHSs$-$} To define a suitable parameter space, we take a leaf from the study of the Heisenberg model.  While any pair of spins in the lattice can interact, the interaction is often short ranged, so only a few nearest neighbors experience significant interactions.  Hence, useful models can be restricted to only a few exchange interactions.  In practice, the analogous procedure is often followed in cuprate studies, where the cuprates are defined by one-band tight-binding models involving only two or three nearest neighbors.  Hence, we take our parameter space as the space of three hopping parameters, $t$, $t'$, and $t''$, with dispersion
\begin{equation}
E=-2t(c_x+c_y)-4t'c_xc_y-2t''(c_{2x}+c_{2y}),
\end{equation}
where $c_{nr} = cos(nk_ra)$, $a$ is the lattice constant, $r = \{x, y\}$, and the VHS crosses the Fermi level when $E_f=E_X=4 (t'-t'')$, where $X =(\pi,0)$.  Since the nearest neighbor hopping only sets an energy scale, we have a 2D parameter space, $t'/t$, $t''/t$.  Cuprates are known to follow a line in this parameter space, with $t''=-0.5t'$, $t'<0$.\cite{PavOK,MBMB,superV}

\begin{figure}[h!]
\centering
\includegraphics[width=0.99\columnwidth]{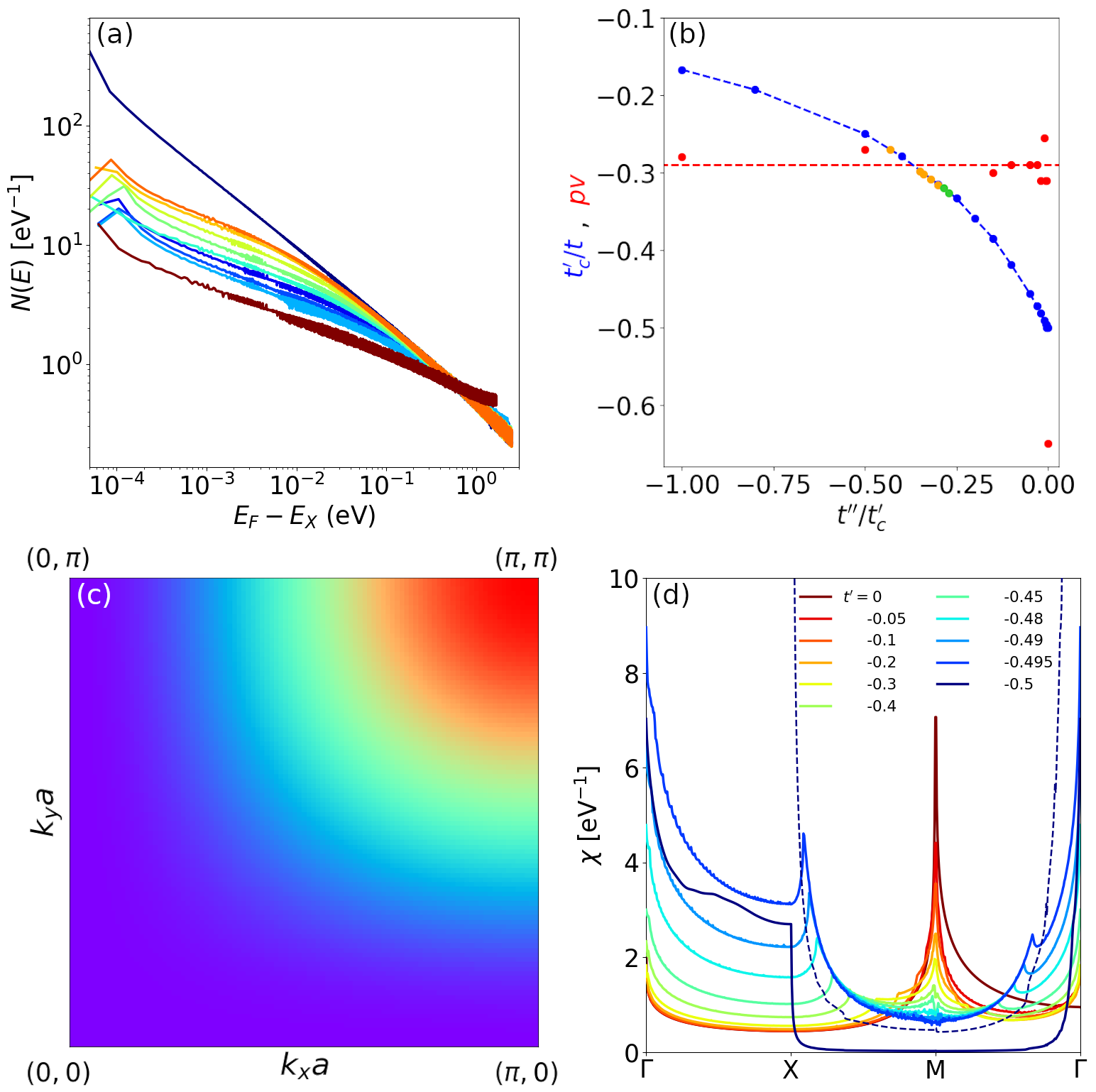}
\caption{(a) {\bf Power-law divergence of $N(E)$ at $t'_c$} for various $t''/t'$ =(from top to bottom)  0.0 , -0.05, -0.10, -0,15, -0.20, -0.25, -0.258 -0.30, -0.35, -0.40, -0.45, and -0.50.  (b) $t'_c$  (blue line) and power-law exponent $p_V$ (red filled circles) vs $t''/t'$.  (c) Dispersion for $t''=0$, $t'=-t/2$.  (d) Susceptibility along high-symmetry directions for $t-t'$ reference family.  For $t'=-0.5t$, the blue dashed curve gives the susceptibility while the solid blue curve is the susceptibility divided by 20. 
}
\label{fig:2Suscept}
\end{figure}

We focus on the parameter range $0\ge t'\ge-t$, $-t'/2\ge t''\ge 0$, which includes the models most often used for cuprates, $t''/t'$ = -0.5, 0.  By fixing $t''/t'$ and varying $t'$, we follow rays emanating from the original Hubbard model, $t'=t''=0$.  For each ray we find a single hoVHS, blue line in Fig.~\ref{fig:2Suscept}(b).  This line has a simple analytic interpretation\cite{superV}, of maximizing the degree to which the fermi surface at the VHS is tangent to the $x$- and $y$- axes -- i.e., maximizing the one-dimensionality.  Thus, for the VHS at $(0,\pi)$, $\partial E/\partial (ak_x) = 2s_x(t+2t'c_y+4t''c_x)\rightarrow 2s_x(t+2t'-4t'')$, where $s_x=sin(k_xa)$.  This vanishes when 
\begin{equation}
t'_c/t = - 1/(2-4t''/t'),
\label{eq:2}
\end{equation}
which is the blue line.  The colored dots on the line are determined numerically by optimizing the linearity of the DOS slope in a log-log plot, as in Fig.~\ref{fig:2Suscept}(a).

\begin{figure}[h!]
\centering 
\includegraphics[width=0.5\textwidth]{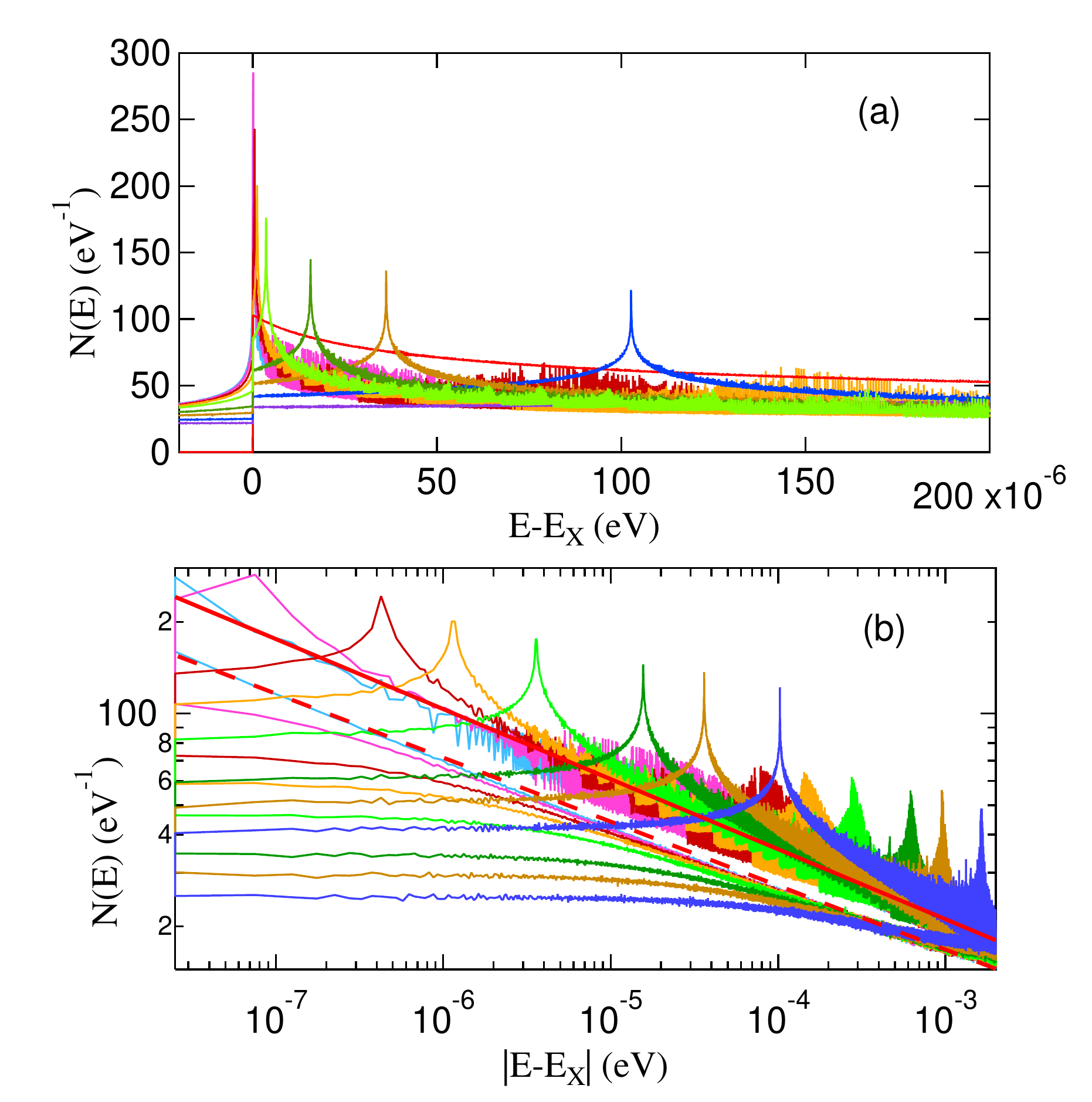}
\caption{{\bf Tuning $t'$ away from $t'_c$ for $t''/t'=-0.002$.}  Frame (b) is a log-log plot of the data of frame (a).  Note that in frame (b), each curve appears twice, the lower one for $E-E_X$ negative.  Both sets of curves display common initial power-law divergences before deviating at low $|E-E_X|$, with different slopes, $p$ = -0.23 (red solid curve) on positive side, -0.21 (red dashed curve) on negative side.}
\label{fig:A1b}
\end{figure}

Figure~\ref{fig:A1b} shows how the DOS evolves as the critical point $t'_c$ is approached.   For $t'<t'_c$, the DOS starts to grow as a power law far from $E_X$, but at some point the curve crosses over to a  logarithmic peak at an energy away from $E_X$. Exactly at $t'_c$, the DOS remains power law to the lowest energies.  Thus, while each ray contains only a single hoVHS, there is a range of parameter space with enhanced logarithmic divergence, which can be much larger than the conventional logarithmic VHS peak.  We note that this pattern of behavior, for $t''=-0.002t'$, is a scaled version of Fig.~1(c) in Ref.~\onlinecite{superV} for cuprates, with $t''=-0.5t'$, and the same behavior is repeated for most of the parameters we have studied.

From data similar to Fig.~\ref{fig:A1b}, we determine the hoVHSs lying in our parameter range, and the results are plotted in Fig.~\ref{fig:2Suscept}.  Figure~\ref{fig:2Suscept}(a) summarizes this divergence for a series of ratios $t''/t'$ in the range 0 to -0.5. While the black curve ($t''=0$) has a single power-law divergence with slope $p_V=-0.65$, the other curves start out at low $E-E_X$ with a lower slope which is the same for all these curves, $p_V=-0.29$ (Fig.~\ref{fig:2Suscept}(b)). This numerically confirms the slope quantization of families of higher-order VHSs\cite{Liang3}. However, above a certain $E-E_X$ the curves deviate from the constant slope towards a higher slope.  For low $|t''|$ the curves merge into the $p_V=-0.65$ curve, while for larger $|t''|$ they cross it.  This is discussed further below.

Figures~\ref{fig:2Suscept}(c,d) clarify that the larger power-law exponent for the $t''=0$ family has its origin in an underlying electronic one-dimensionality.  To understand the origin of the one-dimensionality, it is convenient to look at the dispersion of the state with the strongest instability, corresponding to $t''=0$ and $t'=-t/2$ in Fig.~\ref{fig:2Suscept}(c).  For these parameters, $E=-2t[c_y+c_x(1-c_y)]=-2t$ for $k_y=0$ which is independent of $k_x$.  This dispersion is thus flat along the $y$-axis as well as along the $x$-axis due to symmetry.  Despite this, the susceptibility is not uniform along the $y$-axis due to the crossing of the $x$-axis susceptibility at $\Gamma$ (see Fig.~\ref{fig:2Suscept}(d)). 

The susceptibility is largest at $\Gamma=(0,0)$, corresponding to the DOS.   The lineshape of $N(E)$ is extremely asymmetric since the VHS falls at the bottom of the band. Thus the DOS has a step from zero to infinity on one side, and the power-law fall-off on the other side (Fig.~\ref{fig:2Suscept}(a)).  For smaller $|t'|$, the susceptibility decreases rapidly in  Fig.~\ref{fig:2Suscept}(d), and the DOS at the VHS reverts to the conventional logarithmic form expected for 2D electrons.  We note in passing that since the dispersion is flat along the $x$ and $y$-axes, it cannot be represented by any function of the form $f_1(k_x)-f_2(k_y)$. A finite $t''>0$ modulates the dispersion along the axes, greatly weakening the divergence. The strongest residual divergence arises at the point when $(\pi,0)$ pockets first form.   The DOS retains a quasi-1D lineshape with a power-law divergence with a weaker power $p_V$ on one side and a step down on the other side.

As discussed in the Appendix, to avoid artificial broadening we bin the DOS data over small energy windows.  This leads to statistical errors in $N$ at the lowest energies, clearly seen in Fig.~\ref{fig:2Suscept}(a), but this can be corrected by going to smaller bin sizes.  The numerical slope we find $p_V\sim $0.29 is slightly larger than the analytical result 0.25\cite{Zehyer,Liang3,Ref2b}.  In the Appendix, we extend the numerical calculation to smaller values of $|E-E_X|$ and find values of $p_V$, closer to the analytic result, both for $E>E_X$ and $E<E_X$, as predicted\cite{Liang3}.

In Ref.~\onlinecite{superV}, we showed that a hoVHS need not be a point-like object in momentum space, but could be extended over a finite line segment or area.  Our 1D hoVHS, with $t''=0$, $t'=-t/2$, gives a clear example of this.  Here, when $k_y$ is small the dispersion has the form $E = b(k_x)k_y^2$ along the whole $k_x$-axis, leading to a 1D VHS. Moreover, $b\sim 1-cos(k_xa)\rightarrow 0$ as $k_x\rightarrow 0$, leading to a 1D high-order VHS with $p_V=0.65>0.5$, the conventional 1D result. The reason for the discontinuous change of slope as $t''\rightarrow 0$ is explained in the Appendix.

\begin{figure}[h!]
\centering
\includegraphics[width=0.99\columnwidth]{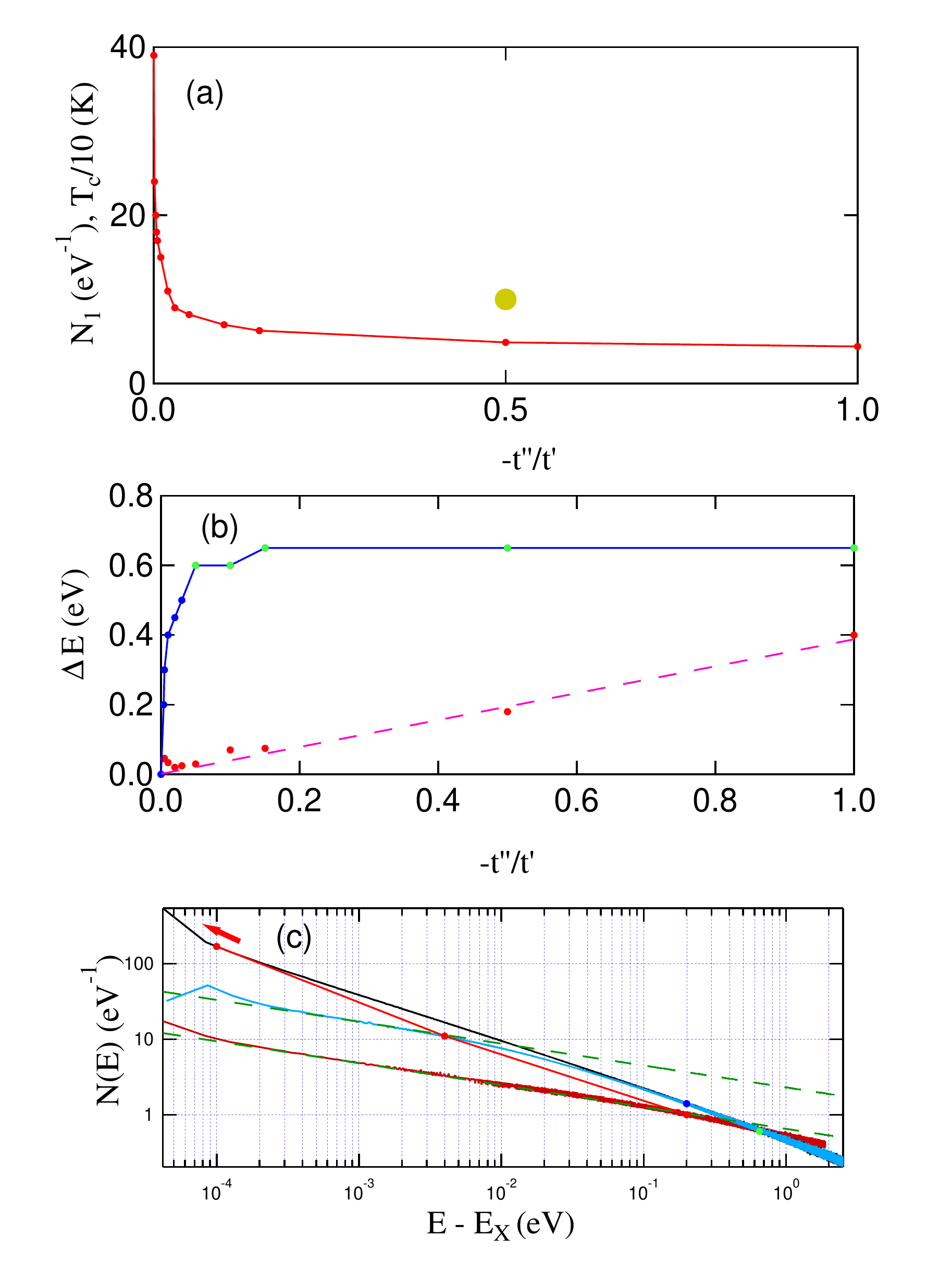}
\caption{(a) {\bf hoVHS DOS at $\delta E$ = 1~meV} vs $t''/t'$.  Gold dot = maximum $T_c$ (left axis) for $t''/t'$. (b) {\bf Crossover behavior from $p_V=-0.29$ to -0.65,}  showing when the DOS starts to deviate from $p_V=-0.29$ (red dots) and when it merges with (blue dots) or crosses (green dots) the $p_V=-0.65$ (green dots) line.  Dashed magenta curve is a straight line through the origin.  (c) Log-log plot of selected DOSs from Fig.~\ref{fig:2Suscept}(a), showing how the data in frame b were calculated -- color of dots is same as in frame b.
}
\label{fig:3b}
\end{figure}


{\it VHS and Superconductivity $-$} 
A closer look at Fig.~\ref{fig:2Suscept}(a) reveals a conundrum.  While all curves with $t''\ne 0$ have the same slope $p_V=0.29$, the amplitude of the DOS increases with decreasing $|t"|$, as the system becomes more 1D.  The cuprates, with $t"=-0.5t’$, have one of the smallest DOSs and are one of the curves that does not merge with the $p_V=0.65$ curve at high energies, Fig.~\ref{fig:3b}(b) (i.e., they are just on the threshold of 1D behavior).  This is more clearly seen in Fig.~\ref{fig:3b}(a), which plots the DOS at $\delta E=1$~meV (the choice of $\delta E$ is arbitrary, as long as all curves with $t''\ne 0$ are on the $p_V=0.29$ branch).  The figure also includes a gold dot for the only known maximum $T_c$ for this group -- the cuprates.  The implication of this figure is manifest: the main phase driven by the hoVHS is not superconductivity, but a phase that competes with superconductivity.  The similarity of this figure to the characteristic phase diagram of exotic superconductors -- forming a dome over the quantum critical point of a competing phase -- is striking, and suggests that fluctuations of the competing order may enhance cuprate superconductivity. However, the cuprates are already known to have a dome near where antiferromagnetism ends, strongly suggesting that two order parameters need to be simultaneously tuned for high-$T_c$ materials to arise.  This is in line with ideas advanced from angle-resolved photoemission studies\cite{Ruihua,Vishik}.  

\begin{figure}[h!]
\centering 
\includegraphics[width=0.56\textwidth]{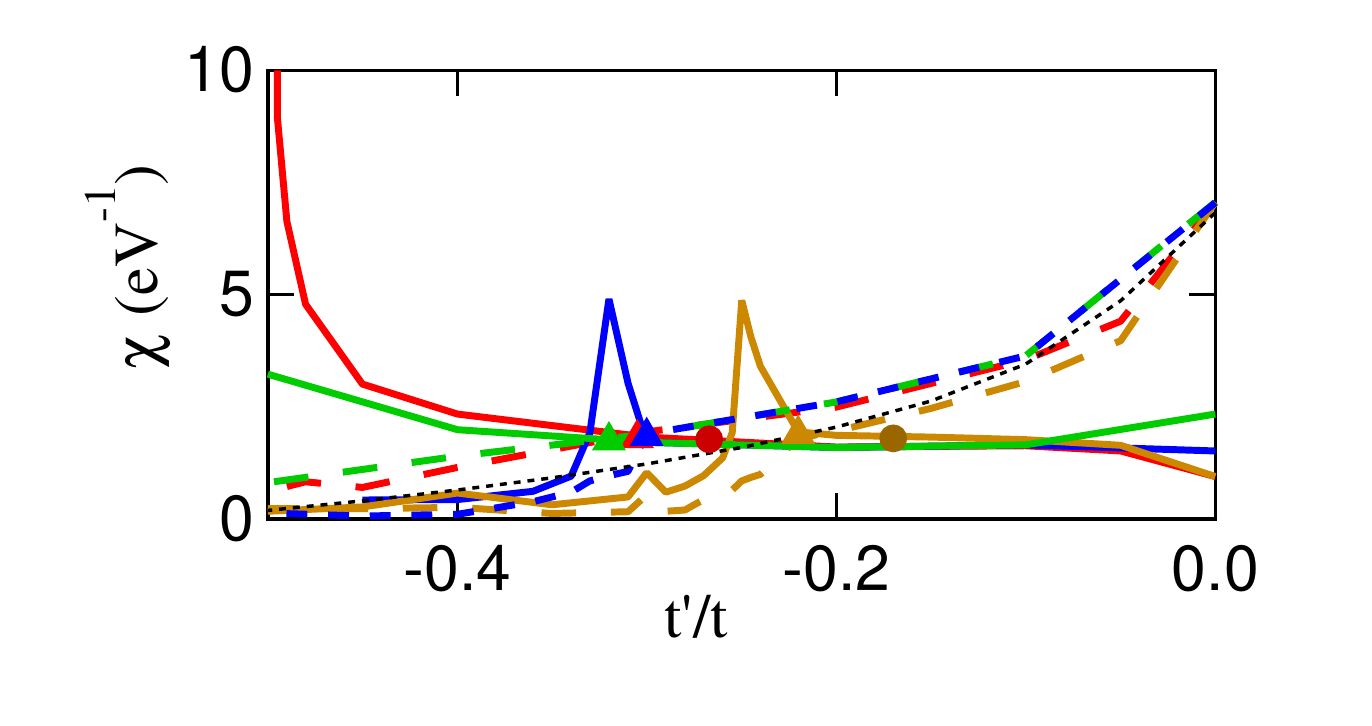}
\caption{ Competition between near-$(\pi,\pi)$ (dashed lines) and $\Gamma$-VHSs (solid lines), for $t''/t'$ = 0 (red lines), -0.25 (blue lines), -0.5 (brown lines), or -1 (green lines).   Corresponding colored triangles indicate crossover from $(\pi,\pi)$ to $\Gamma$-dominated VHS.  For comparison, dark red ($t''=0$) and dark brown ($t''=-t'/2$) circles indicate the positions of the corresponding $x=0$ commensurate-incommensurate transitions discussed in Ref.~\onlinecite{MBMB}.}
\label{fig:3} 
\end{figure}

There are experimental clues in the cuprates that the competing phase may be associated with CDW order.  In line with this, it is found that the cuprate susceptibility diverges at two distinct momenta, $q=0$ -- i.e., the DOS -- and $q=(\pi,\pi)$, related to the AFM order.  These have a distinct competition as a function of either doping $x$ or $t'$.\cite{superV}  In Fig.~\ref{fig:3}, we show that this competition between $(\pi,\pi)$ nesting and $\Gamma$-nesting persists over the full parameter range studied.  In all cases, the dominant susceptibility peak shifts from $q=(\pi,\pi)$ to $q=0$ as $|t'|$ is increased.  Notably, the $(\pi,\pi)$ susceptibility is only weakly dependent on $t''$: the dotted black line in Fig.~\ref{fig:3} is proportional to the analytic approximation\cite{superV} $\chi((\pi,\pi),0)\sim log(T)log(T_X),$ with $T_X=max\{T,T_{eh}\}$ and $k_BT_{eh}=|t'|$ , and provides a good approximation for a wide range of $t''$ values.  For each pair of curves in Fig.~\ref{fig:3}, $t'_{cross}$, the crossover from $(\pi,\pi)$-dominated to $\Gamma$-dominated susceptibility, is denoted by a triangle of the same color.  These tend to cluster near $t'=-0.3t$, but can be pinned by a nearby hoVHS, as when $t''=-0.5t'$, brown triangle.  These crossovers closely scale with the commensurate-incommensurate [$(\pi,\pi)-(\pi,\pi-\delta)$] transition in the undoped cuprates, colored circles, which has been identified as the Mott-Slater transition.\cite{MBMB}  Similar effects arise with doping, {\it i.e.} for $x \ne x_{VHS}$, leading to an $x_{cross}$ with properties similar to $t'_{cross}$.

However, there are also problems with the CDW interpretation.  Thus, experiments find that the CDW phase onsets at a lower temperature than the pseudogap and terminates at a distinctly lower doping than the pseudogap, suggesting that the CDW cannot be the competing order that controls the pseudogap.  Moreover, first-principles DFT calculations in YBCO$_7$ find that the low-energy states continue to be controlled by large magnetic moments associated with fluctuating AFM or stripe order\cite{YUBOI}.  Indeed, low-$q$ or $q\rightarrow 0$ instabilities need not be associated with CDWs, but can include spin-density wave, ferromagnetic, or nanoscale charge inhomogeneity.  Thus, if we could find materials corresponding to values of $t''/t'$ to the left of the cuprates in Fig.~\ref{fig:3b}(a), we could better understand the nature of this second competing phase in cuprates.  Identifying this phase could have implications for the mechanism of superconductivity.  
Indeed, in most cuprates superconductivity terminates near the pseudogap collapse, so optimal superconductivity must fall close to $x_{cross}$.  That makes sense because an electron-electron driven instability such as superconductivity is at a disadvantage compared to an electron-hole instability such as AFM or CDW.  However, when two e-h instabilities are competing, superconductivity can tilt the balance, acting as a symbiotic parasite.  Also, fluctuations will be large near $x_{cross}$ and can further enhance $T_c$.\cite{ABB}

{\it Improving data mining for superconductivity --}


A recent machine-learning study explored the use of DOS peaks to predict the presence of a superconducting instability \cite{CavaEmptor} and found a strong correlation between VHSs and superconductivity for hexagonal but not for square lattices. Our results confirm the latter finding, but ofer a means to improve the correlations.


In Fig.~\ref{fig:3b}(b), we analyze two characteristic features of the DOS lineshapes at the hoVHS: when they start to deviate from $p_V=-.29$ power law behavior (red dots) and when they either merge (blue dots) with or cross (green dots) the $p_V=-0.65$ curve.  The latter are not probative since the crossing point is insensitive to $t''$.  On the other hand, the deviation energy from the 0.29 power law scales linearly with $t''/t'$, making it a useful probe of dimensional reduction.  Figure~\ref{fig:3b}(c) illustrates how these characteristic features are determined.  

\begin{figure}[h!]
\centering 
\includegraphics[width=0.56\textwidth]{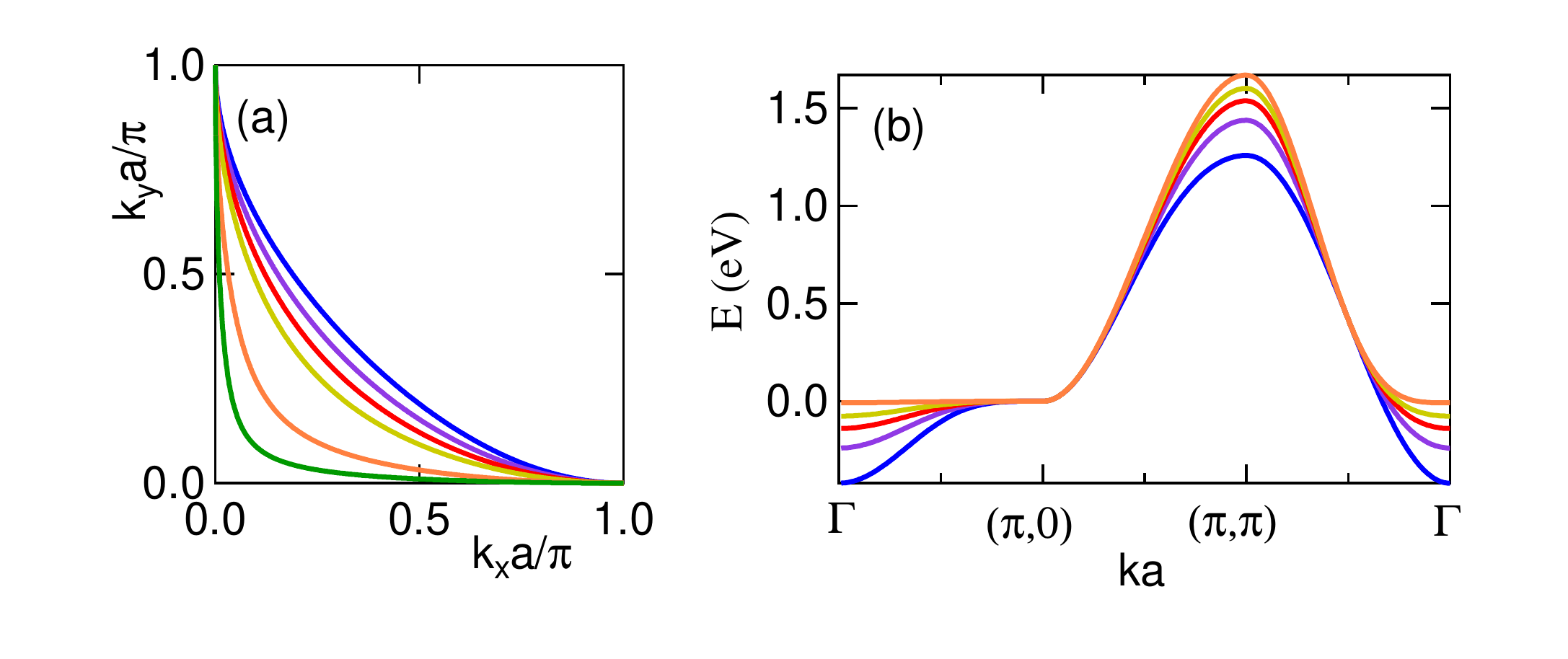}
\caption{Fermi surfaces (a) and dispersions (b) at a series of hoVHSs, calculated using Eq.~\ref{eq:2}, with $t''/t'$ = -0.5 (blue), -0.2 (violet), -0.1(red), -0.05 (gold), -0.005 (orange), and -0.0005 (green).}
\label{fig:5} 
\end{figure}

Alternatively, when band dispersions and fermi surfces are available, these can be used to find the hoVHSs,as illstrated in Fig.~\ref{fig:5}.

{\it Summary and Conclusions$-$}
We have elucidated the origins of the correlation between hoVHSs and high-$T_c$ superconductivity in cuprates\cite{superV} by exploring a wide parameter range away from the cuprates.  We find two kinds of hoVHS associated with one-dimensional electrons -- a strong 1D VHS when $t''=0$ and a weaker version associated with $t''>0$.  Notably, cuprates belong to this latter class, but retain only weak signs of 1D behavior.  In this sense, superconductivity requires just the right amount of hoVHS. Combining this result with earlier known results, high-$T_c$ superconductivity requires simultaneous tuning of two competing orders, perhaps explaining why high-$T_c$ superconductors are so rare.  These results are in line with earlier findings of a clear correlation that cuprates with stronger VHSs have higher $T_c$, even though $T_c$ does not optimize when the VHS is at the Fermi level.

Our study suggests an approach for gaining a deeper understanding of hoVHSs and their role in driving superconducting and other phase transitions.  Using DFT or photoemission data, it is straightforward to extract the three tight-binding parameters used in our analysis for any square-lattice material.  In particular, nearly localized $d$ and $f$ electron systems typically require just a few hopping parameters to allow building up a collection of materials that lie on or near the hoVHS line.  [Note that if signs of both $t'$ and $t''$ are reversed, the same hoVHS line will describe electron instead of hole doping.] The same approach can be readily extended to materials with other symmetries.

The $t-t'-t''$ reference families are convenient for rapidly mapping parameter space and finding the approximate locations of all hoVHSs. However, we anticipate that more distant hopping terms can play the role of `dangerously irrelevant' variables, particularly when one is trying to tune close to a stronger hoVHS, such as the $p_V=0.65$ state.  By relating hoVHSs to catastrophe theory folds, Ref.\onlinecite{Ref2b} provides an informative example of how $k\cdot p$ expansions can lead to the wrong hoVHS if too few terms are retained in the expansion.


\appendix

\section{Appendix: DOS calculations}
 To determine the nature of the DOS divergence at the Van Hove singularity (VHS), the DOS calculations are carried out at $T=0$ to mimic the absence of disorder and nanoscale phase separation. Since the DOS is defined as a sum of delta-functions, we use a binning technique to minimize artificial broadening,
$$N(E) =\frac{2}{N_0}\sum_k\delta (E-\epsilon_k)\simeq\frac{2}{N_0}\sum_iN_i\delta(E-E_i),$$
where $N_0$ is the total number of $k$-points and the factor of 2 is used for the spin degeneracy. $N_i=\int_{E_i-\Delta/2}^{E_i+\Delta/2}N(E)dE$ is the number of $k$-states with energies $E_i=(i-1/2)\Delta$ in the $i^{th}$ bin.  Thus, as $\Delta\rightarrow 0$ and $N_0\rightarrow\infty$, $N_i(E=E_i)\rightarrow N(E)$.  We write $N_0 = (2N_k+1)^2$, where $N_k$ is the number of $k$-values along the positive $x-$ or $y-$axis. 

\begin{figure}[h!]
\centering 
\includegraphics[width=0.5\textwidth]{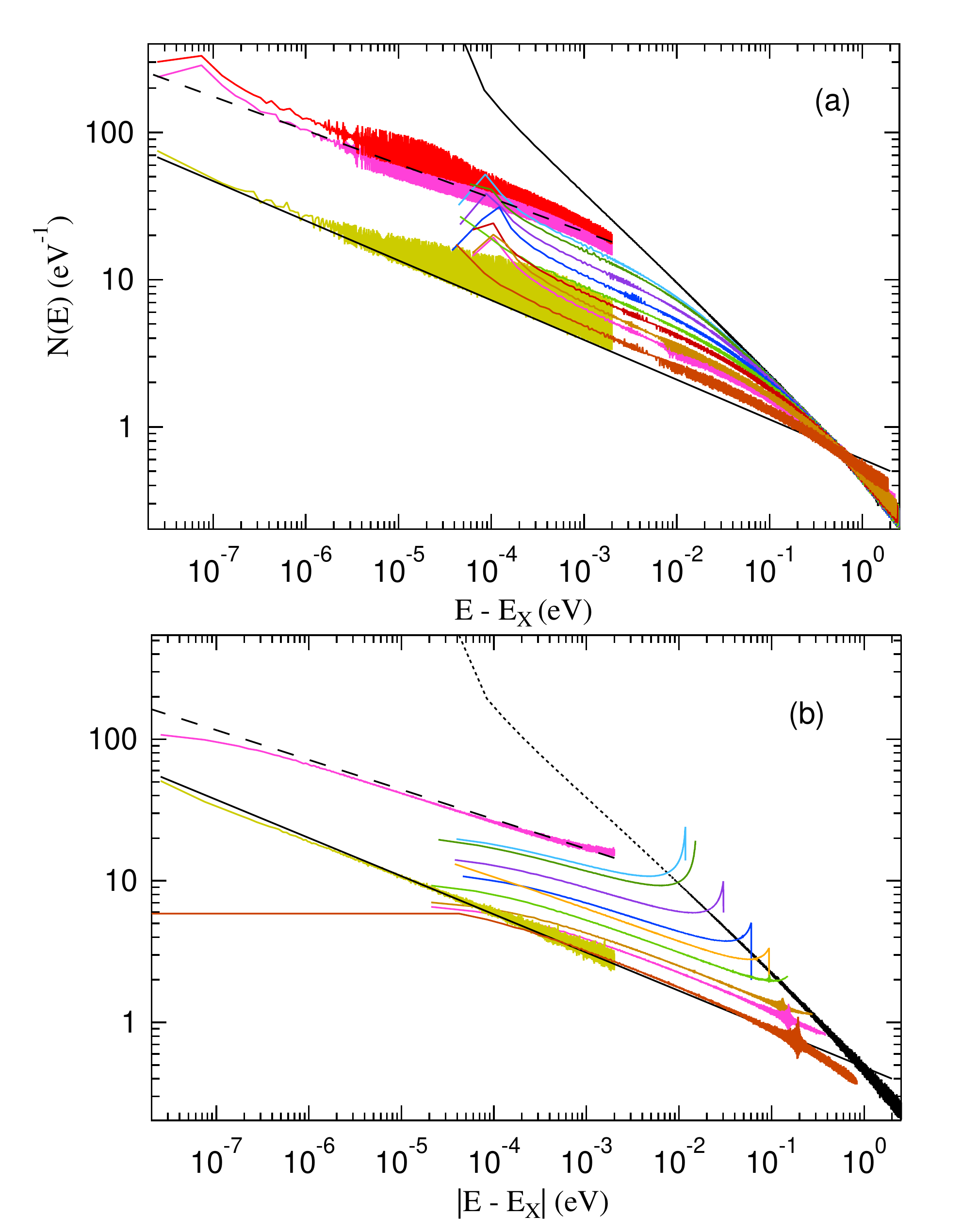}
\caption{{\bf Power-law divergence of $N(E)$ at $t'_c$.}  (a) N(E) at $t'_c$ for various $t''/t'$ values as shown in Fig. 1(a) but with added higher-resolution data for $E-E_X < 2$ meV: $t''/t'$ = -0.5 (gold), -0.003 (magenta), -0.001 (red).  Black lines = power law fits, with $p_V$ = 0.27 (solid line), 0.23 (dashed line). (b) Same as (a) but for the singularity appearing at $ E-E_X < 0 $.  Black lines = power law fits, with $p_V$ = 0.27 (solid line), 0.21 (dashed line).}
\label{fig:A1c}  
\end{figure}

We present DOS in Fig.~\ref{fig:A1c}(a) at finer energy scale than in Fig.~2(a) using a smaller bin size with a denser $k$-mesh. The high-order VHSs are clearly resolved.  At this energy scale, the exponent $p_V$ appears slightly smaller.  It should be noted that the increased broadening of the curves is a sign of statistical error due to under $k$-point sampling. The problem is least in the range of small $\delta E$s, where $N(E)$ is largest and is unimportant for larger $\delta E$s, where larger bins can be used. 

We note an unusual feature in Fig.~\ref{fig:A1c}(b).  The 1D DOS corresponding to $t''=0$ exists only for $E_F-E_X > 0$, whereas the DOS for finite $t''$ is present on both sides of $E_X$.  The black line in Fig.~\ref{fig:A1c}(b) represents the mirror image of the $t''=0$ DOS.  It can be seen that the finite-$t''$ DOS below $E_X$ extends slightly beyond this mirror image, and that the DOS in the interval beyond the mirror image is enhanced, terminating in a finite peak at the band edge.

Finally, we ask how does the slope discontinuity arise as $t''\rightarrow 0$?  As $t''$ decreases, $E_X$ moves toward the top of the band at $E_T$, and part of the DOS intensity shifts to $E_T$, Fig.~\ref{fig:A1c}(b).  While the peak intensity at $E_T$ remains finite for $t''\ne 0$, it tends to diverge as $t''\rightarrow 0$, and the merger of this peak with the higher-order VHS at $E_X$ as $t''\rightarrow 0$ leads to the creation of a new kind of higher-order VHS with larger $p_V$.

\section*{Acknowledgements}     
This work is supported by the US Department of Energy, Office of Science, Basic Energy Sciences grant number DE-FG02-07ER46352, and benefited from Northeastern University's Advanced Scientific Computation Center (ASCC) and the allocation of supercomputer time at NERSC through grant number DE-AC02-05CH11231.  The work at LANL was supported by the U.S. DOE NNSA under Cont. No. 89233218CNA000001 through the LANL LDRD Program and the CINT, a DOE BES user facility.  We thank Adrian Feiguin for stimulating discussions. 

\section*{Author contributions}
R.S.M., B.S., C.L., and A.B. all contributed to the research reported in this study and the writing of the manuscript.

\section*{Additional information}
The authors declare no competing financial interests.

\end{document}